\documentclass{elsart}

\usepackage{natbib}

\usepackage{graphicx}

\usepackage{amssymb}

\def\nutau{$\nu_{\tau}$}

\def\gtrsim{\mathrel{\hbox{\rlap{\hbox{\lower4pt\hbox{$\sim$}}}\hbox{$>$}}}}
\def\lesssim{\mathrel{\hbox{\rlap{\hbox{\lower4pt\hbox{$\sim$}}}\hbox{$<$}}}}

\begin{document}

\begin{frontmatter}




\title{Rates of Horizontal Tau Air-Showers observable by satellites}

 \author[label1]{D. Fargion}
 \author[label1]{, M. De Santis}
 \author[label1]{, M. Grossi}
 \author[label1]{and P.G. de Sanctis Lucentini}
\address[label1]{Physics Department , INFN, Universit\'a di Roma  "La Sapienza", Pl. A. Moro
2, 00185, Rome,  Italy}



\begin{abstract}

  Up-going and Horizontal Tau Air-Showers, UpTaus and HorTaus,  may trace Ultra High Energy Neutrino
  Tau Earth Skimming at the edge of the horizon.
We show that such events even for minimal GZK neutrino fluxes
could be detected by space telescopes such as the EUSO project:
  These Horizontal Tau Showers will track very long fan-like,  multi-finger showers
   whose signature would be revealed by EUSO, OWL experiments.
   Moreover the additional imprint of their $young$ secondaries ($\mu^\pm$ and $\gamma$ bundles with $e^\pm$
   pair flashes)  might allow to disentangle their nature from the old UHECR secondaries in horizontal
   showers. Indeed at large zenith angles ,  the number of $\mu^\pm$ and secondary
$\gamma$'s from  $old$,  Ultra High Energy Cosmic Rays becomes
comparable. On the contrary up-ward muon bundles from UpTaus and
HorTaus may arise within a $young$ shower  with
  a larger gamma-muon ratio ($\sim 10^2$). Such a very characteristic imprint
   maybe observed by Crown detectors on Mountain, planes or
  balloons in space, as well as by $\gamma$ satellites in Space.
  We estimate the UpTaus and HorTaus rate from
  the Earth and  the consequent event rate of
  $\mu^{\pm}$  bundles, whose flux at $92^o -97^o$ exceed the up-going muon flux induced by the atmospheric
  neutrino.
\end{abstract}


\end{frontmatter}


\section{Introduction}


The study of ultra high energy upward and horizontal $\tau$ air
showers produced by $\tau$ neutrino interactions within the Earth
crust has been considered in recent years as an alternative way to
detect high energy neutrinos.
The problem of $\tau$ neutrinos crossing the Earth is indeed quite
complicated because of the complex terrestrial neutrino opacity at
different energies and angles of arrival. In addition, several
factors have to be taken into account, such as the amount of
energy transferred in the $\nu_{\tau}$ - $\tau$ lepton conversion,
as well as the $\tau$ energy losses and interaction lengths at
different energies and materials. This makes the estimate of the
links between the input neutrino - output $\tau$ air shower very
difficult. Such a prediction is further complicated by the
existence of a long list of theoretical models for the incoming
neutrino fluxes (GZK neutrinos, Z-burst model flux, $E^{-2}$ flat
spectra, AGN neutrinos, topological defects). Many authors have
investigated this $\nu_\tau$ signature, however the results are
varied, often in contradiction among themselves, and the expected
rates may
range over a few orders of magnitude (Fargion, Aiello, \&
Conversano 1999; Fargion 2002; Bertou et al. 2002; Feng et al.
2002; Bottai \& Giurgola 2003,  Tseng et al. 2003; Bugaev,
Montaruli, \& Sokalski 2004; Fargion et al. 2004;  Jones et al.
2004; Yoshida et al. 2004). So far, the majority of the current
studies on this topic is based on Monte-Carlo simulations
assuming a particular model of the incoming neutrino flux. Some
author have chosen ad hoc the crossing depth of the tau; other
considered ad hoc maximal distance for the tau in flight. Most of
the authors focus on the UpTaus tracks in underground detectors.
In previous works we have presented a very simple analytical and
numerical derivation (as well as its more sophisticated
extensions) which takes into account, for any incoming angle, the
main processes related to the neutrinos and $\tau$ leptons
propagation and the $\tau$ energy losses within the Earth crust
(see Fargion et al. 2004a, Fargion et al. 2004b for details). Our
numerical results are constrained by upper and lower bounds
derived in more simple and tested approximations. We have shown
how the effective volumes and masses are more severely reduced at
high energy and we included as a further constraint the role of
the air dilution at high altitude, where $\tau$ decay and the
consequent air-shower may (or may not) take place. We showed
(Fargion et al. 2004a) that our results give an estimate of the
$\tau$ air-shower event rates that exceeds earliest studies but
they were comparable or even below more recent predictions
(Yoshida et al. 2004). Secondly, we  pointed out that the
consequent $\mu^{\pm}$, $e^{\pm}$, $\gamma$ signature of HorTaus
largely differs from that of horizontal UHECR backgrounds (Fargion
et al. 2004b).\\
In this paper we introduce in the calculation of the number of
events an additional suppression factor related to the altitude at
which air showers are observed. This guarantees the optimal
extension and the largest flux for the shower to be detected at
each observational $h$ altitude. In particular we apply all our
previous results to the calculation of the expected number of
events at different altitudes in the atmosphere, with a particular
attention to the EUSO telescope. EUSO is a space born detector
that will be located at an altitude of about 400 km on the
International Space Station (ISS), aimed at the the detection of
UHECR.
 It consists of a
wide angle UV telescope that will look downwards towards the Earth
atmosphere to detect the fluorescence signals induced by  UHECR
showers in the atmosphere. Its aperture is such to cover a surface
as large as $1.6 \times 10^5$ km, therefore it will encompass
AGASA-HIRES and Auger areas. EUSO is scheduled to be launched in
the 2009. Its ability to observe within the down ward atmosphere
layer the long tracks of horizontal showers may show their
peculiar opening, by geomagnetic fields, into three main fingers
defined by their charges: (muon and electron pairs of opposite
charge bent in opposite arcs  and rectilinear $\gamma$ paths for
tens and hundreds of km distances ).

\section{How to estimate the Earth-Skimming Volume-Mass}

To calculate the effective volume we assume that the neutrino
traversing the Earth is transformed in a tau lepton at a depth
$x$, after having travelled for a distance ($D(\theta) - x$). The
column depth $D(\theta)$ defined as $\int \rho(r) dl$, the
integral of the density $\rho(r)$ of the Earth along the neutrino
path at a given angle $\theta$ is shown in Fig. \ref{l_tau}. The
angle $\theta$ is included between the neutrino arrival direction
and the tangent plane to the earth at the observer location
($\theta = 0^\circ$ corresponds to a beam of neutrinos tangential
to the earth's surface) and it is complementary to the nadir angle
at the same location. The probability for the neutrino with energy
$E_{\nu}$ to survive until a distance ($D(\theta) - x$) is
$e^{-(D(\theta) - x)/L_{\nu}}$, while the probability for the tau
to exit the Earth is $e^{- x/l_{\tau}}$. On the other hand, as we
will show in the next section, the probability for the outcoming
$\tau$ to emerge from the Earth keeping its primary energy
$E_{\tau_i}$ is $e^{- x/L_{\tau (\beta)}}$ (where $e^{- x/L_{\tau
(\beta)}}$ $\ll$ $e^{- x/l_{\tau}}$ at energy $E_{\tau} > 3 \times
10^{17}$ eV). By the interaction length $L_{\nu}$ we mean the
characteristic length for neutrino interaction; as we know its
value may be associated to the inverse of the total
cross-section  $\sigma_{Tot}= \sigma_{CC} + \sigma_{NC}$,
including both charged and neutral current interactions. It is
possible to show that using the $\sigma_{CC}$ in the
$e^{-(D(\theta) - x)/L_{\nu}}$ factor includes most of the
${\nu}_{\tau}$ regeneration along the neutrino trajectory making
simpler the mathematical approach (Fargion et al. 2004a).\\
The effective volume per unit surface is given by
\[ \frac{V_{Tot}(E_{\nu})}{A} = \frac{V_{Tot \, \oplus}(E_{\nu})}{ 2
\pi R^2_{\oplus}}=   
\int^{\frac{\pi}{2}}_{0}
 \int^{D(\theta)}_{0} e^{-%
\frac{D(\theta) -x}{L_{\nu_{{CC}}}(E_\nu)}}  e^{\frac{-x}{%
l_{\tau}(E_{\tau})}}\sin{\theta}\cos{\theta}d\theta dx \]
where  $A$ is any arbitrary surface above the corresponding
effective volume. For instance this expression has been first
estimated for all the Earth. In this case A is just half of the
terrestrial surface, due to the request of selecting only the
upward direction.  Under the assumption that the $x$ depth is independent of $L_{\nu}$ and $%
l_{\tau}$, the above integral becomes:

\[ \frac{V_{Tot}(E_{\tau})}{A}=\left(\frac{l_\tau}{1-\frac{l_\tau}{L_\nu}}%
\right) \times
\int^{\frac{\pi}{2}}_{0} \left(e^{-\frac{D(\theta)}{L_{\nu CC
}(\eta E_{\tau}))}}- e^{-{\frac{D(\theta)}{l_\tau
(E_{\tau})}}}\right)
\sin{\theta}%
\cos{\theta}d\theta  \]

%

where the energy of the neutrino $E_{\nu}$ has been expressed as
a function of $E_{\tau}$ via the introduction of the parameter $\eta = E_{\nu}/E_{\tau_f}$%
, the fraction of energy transferred from the neutrino to the
lepton. At energies greater than $10^{15}$ eV, when all mechanisms
of energy loss are neglected, $\eta = E_{\nu}/E_{\tau_f} =
E_{\nu}/E_{\tau_i} \simeq 1.2$, meaning that the 80\% of the
energy of the incoming neutrino is transferred to the newly born
$\tau$ after the $\nu - N$ scattering (Gandhi 1996, 1998),
(Fargion et. all 2004a)). Once the effective volume is found, we
introduce an effective mass defined as

\begin{equation}
\frac{M_{Tot}}{A}=\rho_{out}\frac{V_{Tot}}{A} \label{eq_M_ltau}
\end{equation}

where $\rho_{out}$ is the density of the outer layer of the Earth crust: $\rho_{out}=1.02$ (water) and $%
2.65$ (rock).\\
 The expression of the effective volume in the most
general case becomes (see Fargion et al 2004 for details)

\begin{equation}
\frac{V_{Tot}(E_{\tau})}{A}=\left(\frac{L_{\tau (\beta)}(E_\tau)}{1-\frac{%
L_{\tau (\beta)}(E_{\tau})}{L_{\nu_{CC}}(\eta E_\tau)}}\right) \times 
\int^{\frac{\pi}{2}}_{0} 
e^{-\frac{D(\theta)}{L_{\nu CC
}(\eta E_{\tau}))}}
\sin{\theta}%
\cos{\theta}d\theta \label{Veff_lbeta}
\end{equation}

where  the interaction length $L_{\tau (\beta)}$ (shown in Fig.
\ref{l_tau} and compared to $l_{\tau }$) guarantees a high energy
outcoming $\tau$ even if outcoming from a thinner Earth crust
(see Fargion et al. 2004 for a more detailed
discussion of $L_{\tau (\beta)}$).\\
We remind that the total neutrino cross section $\sigma _{\nu }$
consists of two main component, the charged current and neutral
current terms, but the $\tau $ production depends only on the
dominant charged current whose role will appear later in the event
rate number estimate. The interaction lengths $L_{\tau \beta }$,
$L_{\nu_{CC} },$ depends on the energy, but one should be careful
on the energy meaning. Here we consider an incoming neutrino with
energy $E_{\nu _{i}}$, a prompt $\tau $ with an energy $E_{\tau
_{i}}$ at its birth place, and a final outgoing $\tau $ escaping
from the Earth with energy $E_{\tau _{f}}$, after some energy
losses inside the crust. The final $\tau $ shower energy, which is
the only observable quantity, is nearly
corresponding to the latter value $E_{\tau _{f}}$ because of the negligible $%
\tau $ energy losses in air. However we must be able to infer
$E_{\tau _{i}}$ and the primary neutrino energy, $E_{\nu}$, to
perform our calculation. The effective volume  resulting from Eq.
\ref{Veff_lbeta} calculated for a detector with a 1 km$^2$
acceptance area is displayed in Fig. \ref{Volume}.

\section{From the GZK neutrino flux to Tau-Air Shower rate for EUSO}

After having introduced the effective volume we can estimate the
outcoming event number rate for EUSO for any given neutrino flux.
The consequent event rate for incoming neutrino fluxes may be
easily derived by:

\begin{equation}
\frac{dN_{ev}}{d\Omega dt}= \left( \int \frac{dN_{\nu}}{dE_{\nu}
d\Omega dA dt } \sigma_{N \nu} (E) dE \right) n \rho_r V_{Tot}
\end{equation}
where $L_{\nu CC} = (\sigma_{N \nu} n)^{-1}$,
 $\Phi_\nu= \frac{d\,N_{\nu}}{d\,E_\nu}E_\nu=5 \cdot
10^{-18}\left( {\frac{E_\nu}{10^{19}eV}} \right)^{-\alpha + 1}
$cm$^{-2}$ sec$^{-1}$ sr$^{-1}$ describes as a flat spectrum
($\alpha = 2$) most of the GZK neutrino flux and as a linearly
increasing spectrum ($\alpha = 1$)  the Z-burst model; $\rho_r$ is
the density of the most external layer (either rock or water). The
assumption on the flux may be changed at will  and the event
number will scale linearly according to the model. \\ In Fig.
\ref{EUSOevent} we show the expected number of event for
EUSO 
where we have included the Earth's atmosphere and we have used
$L_{\tau (\beta)}$, so that we may express the results as a
function of the final $\tau$ lepton energy.\\
%
%
%
As one can see from Fig. \ref{EUSOevent}, at energy $E = 10^{19}$
eV the general expected event rate is given by:

\[ N_{ev}\,= 5\cdot 10^{-18}cm^{-2}s^{-1}sr^{-1}\,
\left({\frac{V_{eff} \rho_r}{L_{\nu CC}}}\right) (2 \pi\,\eta_{Euso} \Delta t)  \]  
\[ \times \left( \frac{\Phi_{\nu} E_{\nu}}{50 eV cm^{-2} s^{-1}
sr^{-1}} \right) \eta^{-\alpha} \left( \frac{ E_{\tau}}{10^{19} \,
eV} \right)^{- \alpha + 1}\]

 Such number of events greatly exceed previous results by at least two orders of magnitude (Bottai et al. 2003)
 but are comparable, but below more recent estimates (Yoshida et al. 2004).

\section{The Tau air showers and secondary $\mu$ differential rates}

The differential number of events is given by
\begin{equation}
\frac{dN_{ev}}{d\Omega dt }= \Phi_\nu \frac{\rho_r \,
V_{eff}}{L_{\nu CC}} \label{neventieq}
\end{equation}

We introduce now a differential expression of the number of events
which allows to calculate the number of events as a function of
the angle $\theta$. We shall introduce a suppression factor that
cares the finite lenght of the horizontal atmosphere (Fargion et
all. 2004a).\\
 We can rewrite the expression of the effective
volume given in Eq. \ref{Veff_lbeta} as a differential volume for
each arrival angle $\theta$:

\begin{equation}
\frac{dV}{ d\theta\, d\phi \, d\Omega \, dA}= \left[
1-e^{\frac{-L_{0}}{c\,\gamma_\tau \tau}}\right]
\times \\
 {l_\tau (E_\tau)}
\frac{
e^{-\frac{D(\theta)}{L_{\nu
CC}}}
}{\left( 1-\frac{l_\tau
(E_\tau)}{L_{\nu CC}}\right)}\sin\theta \, \cos\theta
\end{equation}

and we obtain the following expression for the differential rate of
events

$$
\frac{dN_{ev} E}{dE d \Omega \, d\theta\, d\phi\, dt\, dA}=
\Phi_{\nu_0} \eta^{-\alpha} \left(
\frac{E_\tau}{E_{\nu_0}}\right)^{-\alpha + 1} \rho_r
\\
\left[ 1-e^{\frac{-L_{0}}{c\,\gamma_\tau \tau}}\right]
 \frac{l_\tau (E_\tau)}{L_{\nu CC}}
\frac{e^{-\frac{D(\theta)}{L_{\nu
CC}}}
}{\left( 1-\frac{l_\tau (E_\tau)}{L_{\nu CC}}\right)}\sin\theta \,
\cos\theta
$$

with $\eta=1.2$ and $E_{\nu_0} = 10^{19}$  eV.\\
If we now integrate on the solid angle $d\Omega$ (half side) we
obtain the above formula multiplied by a factor $2\pi$.\\
Given the $\tau$ number of events we can calculate the rates of
$\mu$, $e^\pm$ pairs and $\gamma$, which originates as secondary
particles from the $\tau$ decay. The number of muons is related to
the total number of decaying pions and according to Matthews
(2001)  is given by


\begin{equation}
N_\mu \simeq 3\cdot 10^5 \left( \frac{E_\tau}{PeV}\right)^{0.85}
\end{equation}

\begin{equation}
N_{e^+ e^-} \simeq 2\cdot 10^7 \left( \frac{E_\tau}{PeV}\right)
\end{equation}

\begin{equation}
N_{\gamma} \simeq  10^8 \left( \frac{E_\tau}{PeV}\right)
\end{equation}

and we obtain finally

\begin{equation}
\frac{dN_i^{ev}}{d\theta\, d\phi\, dt\, dA}(E, \theta)=  N_i \cdot
\frac{dN_{\tau}^{ev}\,E_\tau}{ dE \, d\theta\, d\phi\, dt\,
dA}\nonumber \label{Nevent_multiplicity}
\end{equation}


We show in Fig. \ref{tau_mu} the average differential rate of
$\tau$'s and the secondary $\mu^\pm$, $e^{\pm}$ and $\gamma$
bundles from the decay of $\tau$ leptons.  We find that the muon
signal at the horizon, related to Earth skimming tau neutrinos is
above 10$^{-12}$ - 10$^{-11}$ cm$^{-2}$ s$^{-1}$ sr$^{-1}$. One
should notice that the  muonic background produced by atmospheric
neutrinos (CR $\rightarrow \mu^{\pm} \rightarrow \nu_{atm}
\rightarrow \mu^{\pm}$) below the horizon approaches the value
$\Phi_{\mu_{atm}}\simeq 2\cdot 10^{-13}$ cm$^{-2}$ s$^{-1}$
sr$^{-1}$, which is at least one order of magnitude lower than
what we have obtained from our calculation for minimal GZK
$\nu_{\tau}$ fluxes (see Fig. \ref{tau_mu}).

Moreover this muonic shower would have a significant $\gamma$
component, with a high number of  photons - $N_{\gamma}/N_{\mu}
\sim 10^2$ (Cillis \& Sciutto 2001; see also Fig. 15 in Cronin
2004)
 - because of the $\tau$ decay channels into both charged and neutral
pions.\\
On the other hand, horizontal UHECRs will not represent a source
of contamination for our signal because the horizontal $\gamma$'s
produced in the hadronic shower inside the atmosphere would be
exponentially suppressed at large slant depth ($X_{max} > 3000$ g
cm$^{-2}$) and large zenith angles ($\theta > 70^\circ$) (Cillis
\& Sciutto 2001). Only $\mu^\pm$ can survive when propagating
through the atmosphere at large zenith angles. Such muons would
also be source of parasite $\gamma$ signal - due to the $e^\pm$
pair produced in the $\mu^\pm$ decay in flight - but the
gamma-to-muon-number ratio would be now approximately $\lesssim$
1.\\
Therefore, this difference would allow to distinguish gamma-rich
HorTaus from common horizonal gamma-poor UHECR events to reveal
UHE earth-skimming $\nu_{\tau}$'s .\\
For a more precise approach to the calculation of the rate of
events one has also to take into account that the number of events
varies as a function of the height $h$ at which the observer is
located to detect the muonic, electronic and gamma shower. Therefore
we
introduce 
an additional factor

\[ \frac{dN_{\mu}^{ev}}{  d\theta\, d\phi\, dt\, dA}(E, \theta, h)=
\left( 1 - e^{-\frac{h}{h_s}} \right) \frac{dN_{\mu}^{ev}}{
d\theta\, d\phi\, dt\, dA}(E, \theta) \]

where the parameter $h_s$ that we have introduced as

\begin{equation}
h_s=R_{\tau} (E_{\tau}) \sin{\theta} +
\frac{X_{Max}(E_\tau)}{\rho_r}\sin{\theta}
\end{equation}


defines the optimal height where the shower can reach its maximal
extension at the corresponding energy $E_{\tau}$.  This is the sum
of the height reached by the $\tau$ in the atmosphere before its
decay ($R_{\tau} \sin \theta$), where we have neglected $\tau$
energy losses in the atmosphere, and the altitude reached by the
secondary particles of the shower which is related to  the
parameter $X_{max}$. Note that $R_{\tau} = 4.9 (E_\tau/10^{17} \,
eV)$ km and $X_{max}/\rho_r = 5.7 + 0.46 \ln (E_{\tau}/10^{17}
eV)$ km. Here we have considered the air density $\rho_r = 1.25
\times 10^{-3}$ g cm$^{-3}$ constant and equal to the value at the
sea level. At low energies ($10^{15}$ - $10^{16}$ eV) the second
term is dominant and $\rho_r$ can be considered as constant
because the $\tau$ lepton travels for less than 1 km before it
decays. At higher energy ($10^{17}$ - $10^{19}$ eV) the first term
is dominant, and we can neglect the way the exact value of
$\rho_r$ changes with the altitude. In Fig. \ref{differential} we
show the differential number flux (per unit area, energy and
time) at different altitudes of $\mu$ leptons originated by
$\tau$ decay. We have performed this calculation assuming an
input GZK neutrino flux and that the Earth outer layer is made of
rock (left panel) and water (right panel). We show how the
expected number of events increase with the altitude of the
observer. In particular at $h =$ 40-20 km, roughly the altitude of
the atmosphere layer where HAS or HORTAUs are expanding to
maximal power and where EUSO is following the UHECR showers, the
differential number flux of muons and electromagnetic showers reach already its maximal asymptotic value.\\
In Fig. \ref{MU_FLUX17} we display the differential number flux
(per unit area, solid angle and time) of  the secondary $\mu^\pm$
pair (from HORTAUs) as a function of the aperture $\theta$ which
describes the line of sight below the horizon (i.e. $\theta =
\theta_{zenith} - 90^{\circ}$). Again we have assumed an input GZK
neutrino flux and an Earth outer layer made of rock. The two
panels correspond to two different $E_\tau$ energies: $E_{\tau} =
10^{17}$ ev (Fig. \ref{MU_FLUX17}, left panel) and $E_{\tau} =
10^{18}$ ev (Fig. \ref{MU_FLUX17}, right panel). Again we show
that at higher altitudes the muon, (as well as the gamma flux)
from HORTAUs is high and that satellites orbiting around the
Earth at a few hundreds of kilometers may better search  Earth
skimming  tau-air showers secondaries, if they will turn their
focus on the Horizons edges of the Earth, where they are mostly
produced (See Fargion et all. 2004c)

\section{Conclusions}

Horizontal showers from normal hadrons (or gammas) are strongly
depleted of their electromagnetic component because of the large
slant depth ($X_{max} > 3 \times 10^4$ g cm$^{-2}$), while
horizontal tau air showers are not. Indeed "young" HorTaus either
of hadronic (67\%) or electromagnetic (33\%) nature  at their peak
shower activity are expected  to have a large $N_{\gamma}/N_{\mu}$
ratio, greater than $10^2$ (but with a characteristic energy ratio
$E_{\gamma}/E_{\mu} < 10^2$). Old horizontal showers would have
$N_{\gamma}/N_{\mu} \simeq 1$. This difference would allow to
distinguish and disentangle HorTaus from horizontal UHECR events,
even in absence of good angular resolution, opening a new
perspective in the UHE neutrino Astronomy. The secondary fluxes of
muons and gamma bundles made by incoming GZK neutrino fluxes and
their HorTau showers, is well above the noise (by one-two orders
of magnitude) made by up-going atmospheric muons
The neutrino signals at energies
much above EeV  may be even better probing the expected harder
neutrino Z-Burst model spectra \cite{Fargion-Mele-Salis99},
\cite{Weiler99}.\\
The peak fluences we find  in the $\mu$ and $\gamma$ component at
the horizon ($\pm 5^\circ$) will give a signal well above the
background produced by atmospheric  $\nu$'s. We have not discussed
the albedo muons whose fluxes measured  by Nemo-Decor experiments,
$\phi_{\mu} \lesssim 10^{-9}$ cm$^{-2}$ s$^{-1}$ sr$^{-1}$, are
made mostly by single tracks \cite{decor}. Because pair (or
triple) bundle muons are much
 rarer ($\phi_{\mu_{pair}} < 10^{-4} \phi_{\mu_{single}} \approx 10^{-13}$ cm$^{-2}$ s$^{-1}$
 sr$^{-1}$), the search and detection of muon bundles by GZK
 HorTaus at a minimal rate of $10^{-12}$ cm$^{-2}$ s$^{-1}$
 sr$^{-1}$ (over an area of $10^2$ m$^2$) will lead, in a year, to
 about 30 muons possibly clustered in five - ten multiple bundles.
 These events will be reinforced by hundreds or thousands of
 associated collinear gamma flashes.
A detector with an area of few tens or hundreds of square meters
pointing to the horizon from the top of a mountain would be able
to reveal the GZK $\nu_{\tau} - \tau$ young showers (Iori, Sergi
\& Fargion 2004). The characteristics of a prototype twin
crown-like array detector to be placed on mountains, balloons, or
satellites (Fargion2001a) will be discussed in detail (Fargion
et. all. 2004c)). The simultaneous sharp $\gamma$ bundle at
$\phi_{\gamma} \sim 10^{-9} \div 10^{-11} $ cm$^{-2}$ s$^{-1}$
sr$^{-1}$ and the "burst" of electron pair at $\phi_{e^+e^-} \sim
10^{-9} $ cm$^{-2}$ s$^{-1}$ sr$^{-1}$ would give evidence of
unequivocal $\tau$ signature.\\It should be reminded that the
neutrino interaction enhancement by TeV new Physics
 would produce also an increase of hundreds or thousands time in HorTaus beyond a
mountain Chain (like Auger) than standard weak interactions would
do. Therefore Auger (a)must soon detect the $Andes$ $Shower$  $
Shadows$ toward  the far away west side because of their
absorption inside the mountain;(b) Auger must  reveal the absence
(or) the birth of $young$ (Fargion,Aiello Conversano 1999),
(Bertou et all. 2002), HorTaus  created by tau born inside the
the mountain themselves; they will be more abundant if  New
Physics at TeV is inducing  larger neutrino-nucleon interactions.
In a few years Auger might anyway be  able to observe the
expected GZK neutrinos inducing HorTaus at EeV energies.
 In conclusion we showed that an orbiting telescope such as EUSO experiment will be able to see at
least half a dozen of events of HorTaus mainly enhanced along the
Continental Shelves or Mountain edges.
 To conclude we want to remind that inclined-vertical PeVs
$\tau$ air showers (UpTaus) would nearly always be source of
$\gamma$ "burst" surviving the atmosphere opacity. These sharp
UpTaus (with their companion HorTaus above and near EeVs) might be
observed by satellites as brief Terrestrial Gamma Flashes (TGF).
Indeed we identified a possible trace of such events in BATSE
record (taken during the last decade) of $78$ upgoing TGF possibly
associated with galactic and extragalactic UHE neutrino sources
(Fargion 2002).

\begin{figure}[t]
\begin{center}
\includegraphics[height=50mm,width=13.8cm]{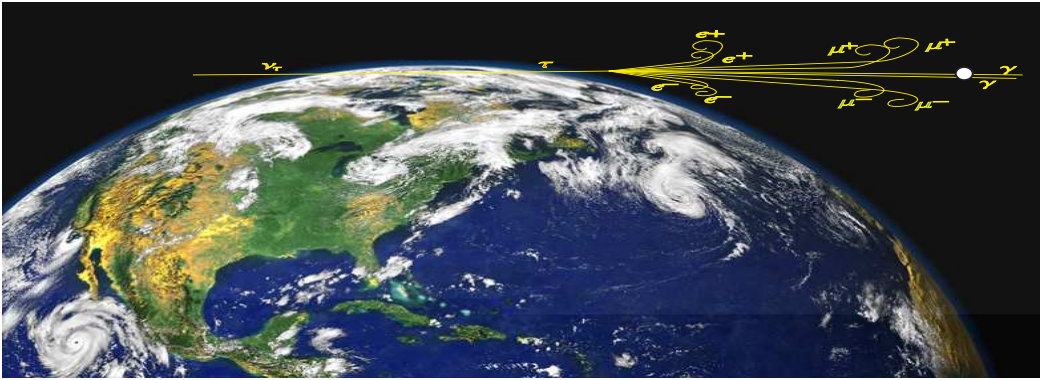}
\caption{Horizontal Upward Tau Air-Shower (HorTauS) originated by
UHE neutrino skimming the Earth: fan-shaped jets arise because of
the geo-magnetic bending of charged particles at high quota ($\sim
23-40$ km). The shower signature may be observable by EUSO just
above the horizon. Because of the Earth
opacity most of the UpTau events at angles $%
\protect\theta > 45-50^o$ will not be observable, since they will
not be contained within its current field of view (FOV).
} \label{fig3}
\end{center}
\end{figure}

\begin{figure}[t] 
\begin{center}
\includegraphics[width=55mm,height=67mm,angle=270]{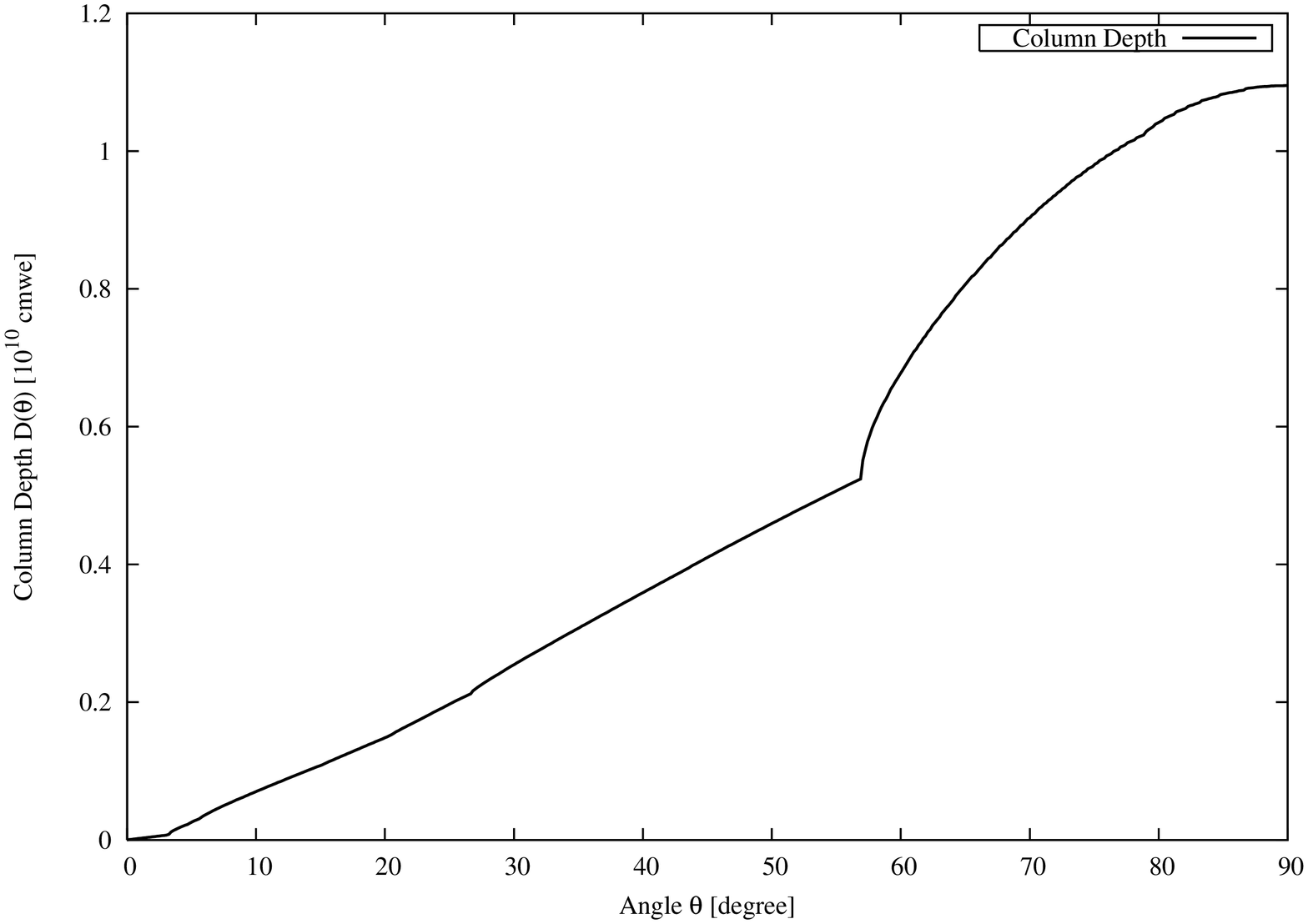}
\includegraphics[width=55mm,height=67mm,angle=270]{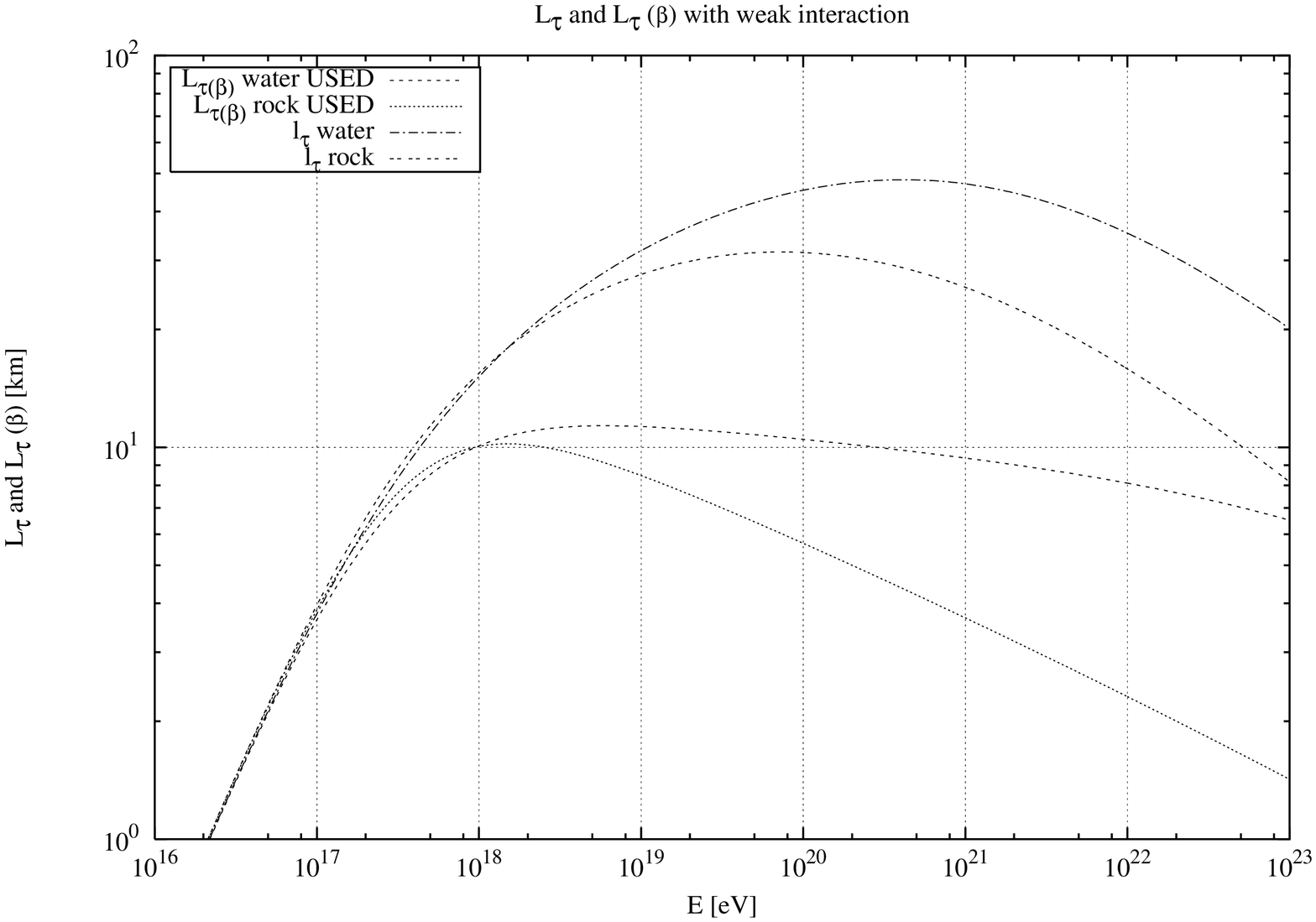} \caption{{\em Left:} Column depth as a function of the incoming angle having
assumed the multi-layers structure  given by the Earth Preliminary
Model.{\em Right:} Comparison between $L_{\tau (\beta)}$ and
$l_{\tau}$ for rock and water.  As one can see from the picture,
$L_{\tau (\beta)}$ is shorter than $l_{\tau}$ at energies above
$10^{17}$ eV, thus it
 corresponds to a smaller effective volume where $\protect\tau$'s are produced while
keeping most of the primary neutrino energy. The energy   label
on the x axis refers to the newly born tau for $L_{\tau(\beta)}$:
$E_{\nu_{\tau}}$} \label{l_tau} \end{center}\end{figure}

\begin{figure}[t]
\begin{center}
\includegraphics[width=52mm,angle=270]{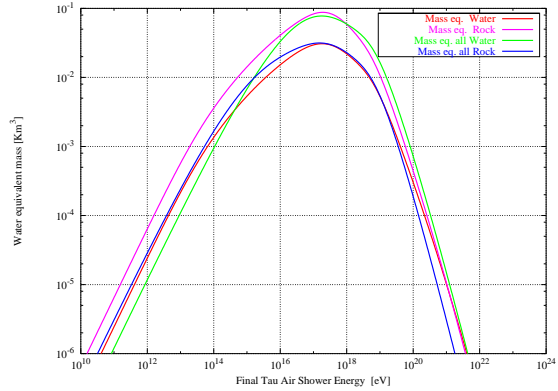}
\caption{Effective mass for UpTaus and HorTaus per km square unit
area including the suppression factor due to the finite extension
of the Earth's atmosphere (an horizontal length  of $600$ km). The
curves obtained (the red line for an Earth outer layer made of
water and the pink line for the rock) are compared with a
simplified model of the Earth, considered as an homogeneous sphere
of water (green line) and rock (blue line). Here we used the
interaction length $L_{\protect\tau (\protect\beta)}$ and the
volume is expressed as a function of the final tau energy. Note
that above $10^{-2}$ km$^2$ the effective mass-volume in the
energy range 3 $\times$ 10$^{15}$ - 10$^{19}$ eV is larger than
the volume of the atmospheric layer, whose ability to convert
downward neutrino in observable air shower is negligible. Only
horizontal neutrino interaction in the atmosphere may be detected,
at a much lower rate than the HorTaus ones.} \label{Volume}
\end{center}\end{figure}

\begin{figure}[t]
\begin{center}
\includegraphics[angle=270,scale=0.3]{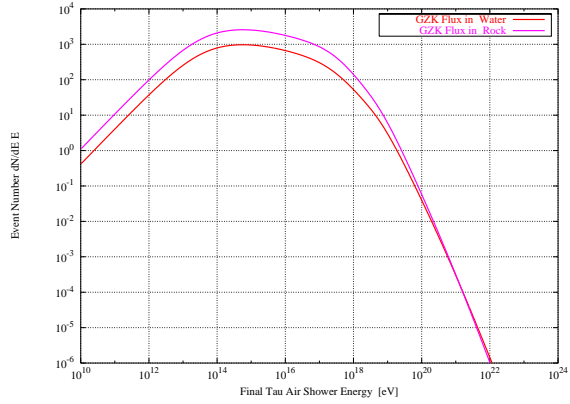}
\caption{Number of EUSO Event for HorTaus in 3 years record as a
function of the outgoing lepton tau ($L_{\tau (\beta)}$ as
interaction length), including the finite extension of the
horizontal atmospheric layer. At energy $E_{\tau} = 10^{19}$ eV,
the event number is $N_{ev}= 3.0$ ($\protect\phi_{\protect\nu}
E_{\nu} / 50$ eV cm$^{-2}$ s$^{-1}$ sr$^{-1}$) for the
water and $N_{ev}= 6.0$ ($\phi_{\protect\nu} E_{\nu} / 50$ eV cm$^{-2}$ s$%
^{-1}$ sr$^{-1}$) for the rock. The resulting number of events has
been calculated for an initial GZK neutrino flux: $\propto E^{-2}$
.} \label{EUSOevent}
\end{center}
\end{figure}

\begin{figure}[t] 
\includegraphics[width=80mm,height=55mm,angle=0]{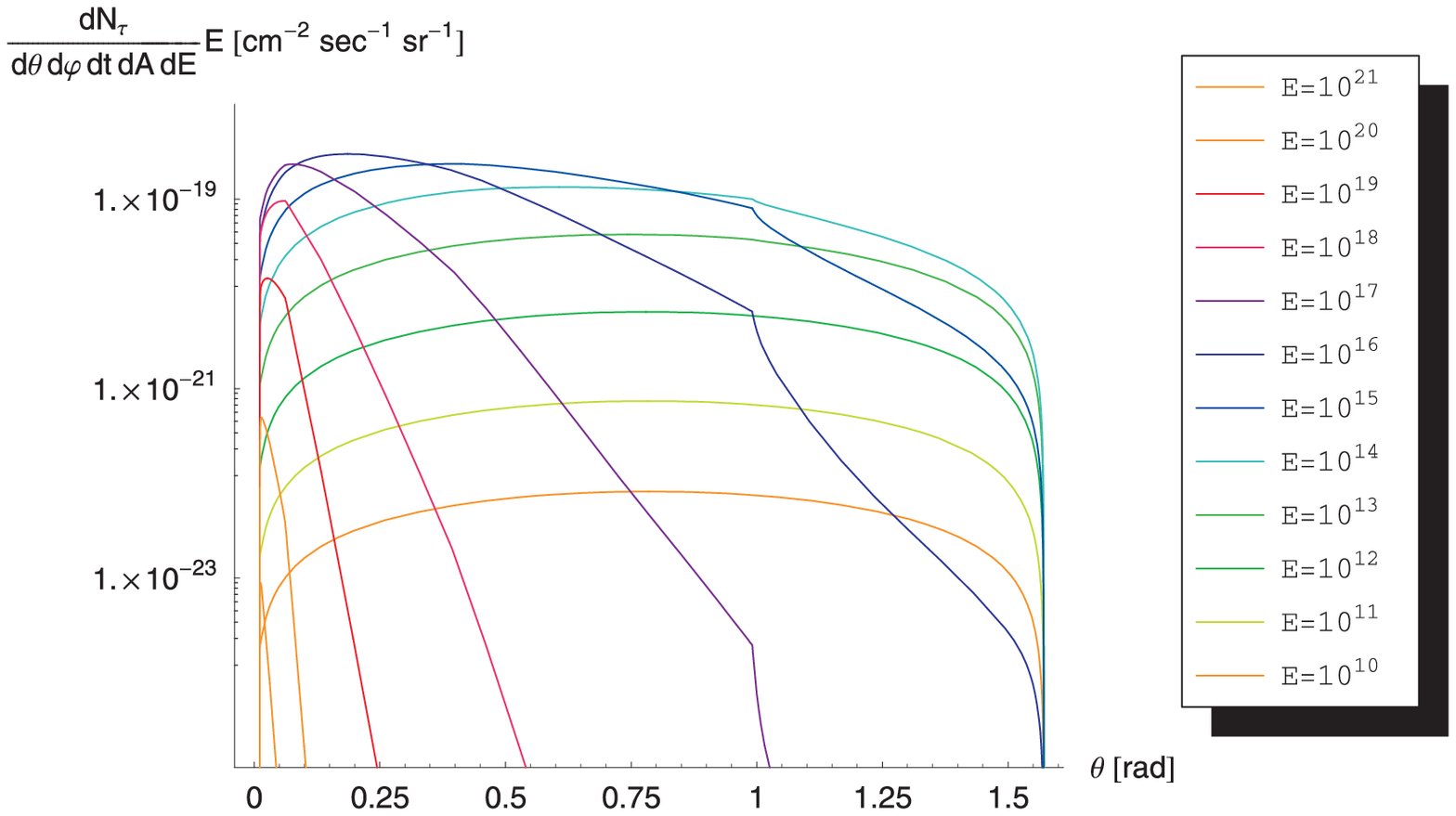}
\includegraphics[width=80mm,height=55mm,angle=0]{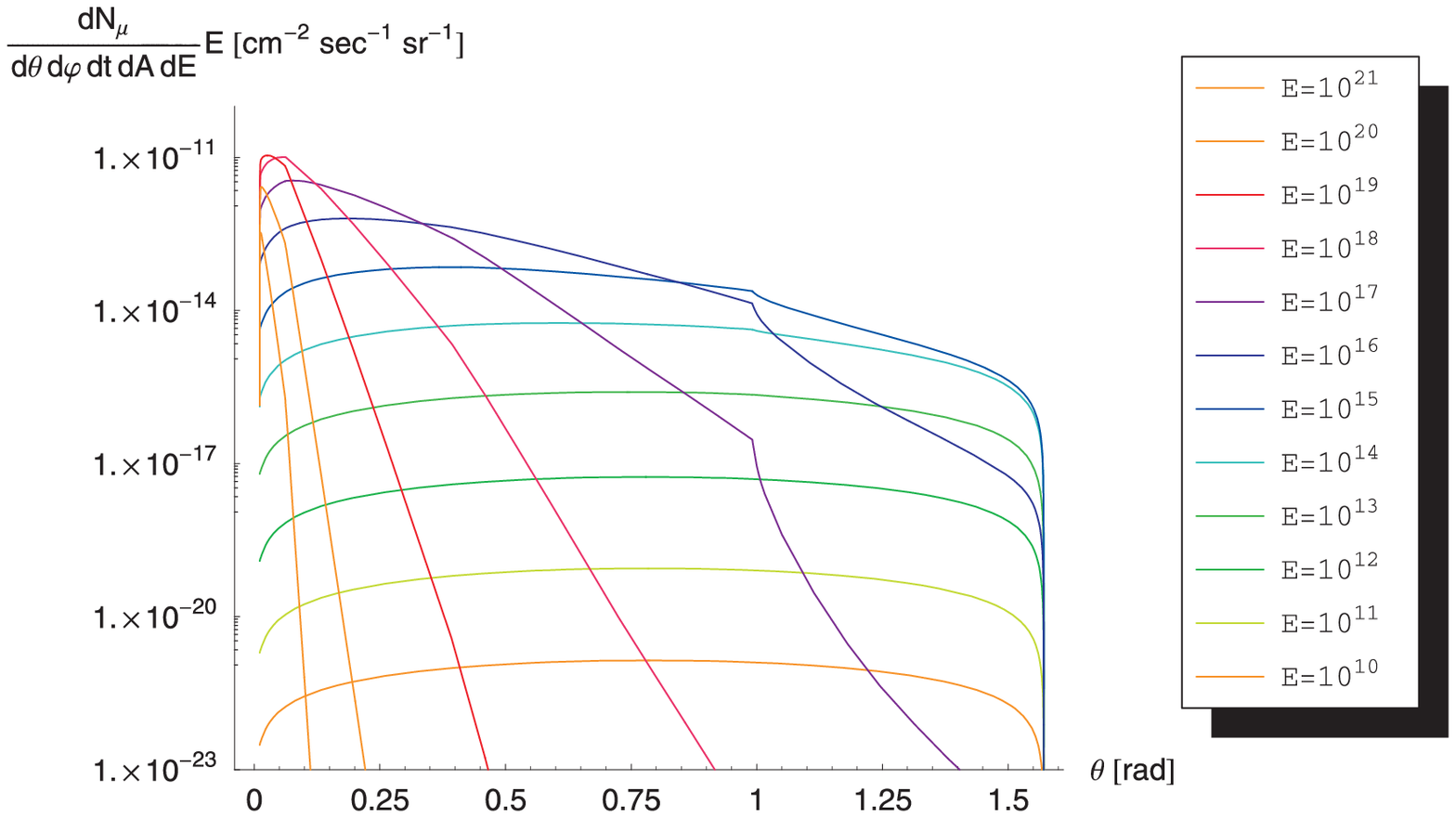}
\caption{{\em Left:}
 The differential  number of event rate of $\tau$
leptons (HorTaus) for an input GZK neutrino flux. As in previous
Figures we are assuming that $\tau$'s are escaping from an Earth
outer layer made of rock. Note the discontinuity at $\theta \simeq
1$ rad, due to the corresponding  inner terrestrial  higher
density core (see Fig. \ref{l_tau}). {\em Right:} The differential
number of event rate of the secondary muons produced by the decay
in flight of $\tau$ leptons in the Earth's atmosphere: HorTaus. As
in previous Figures we are assuming an input GZK neutrino flux and
that $\tau$'s are escaping from an Earth outer layer made of rock.
} \label{tau_mu}
\end{figure}

\begin{figure}[t] 
\includegraphics[width=80mm,height=55mm,angle=0]{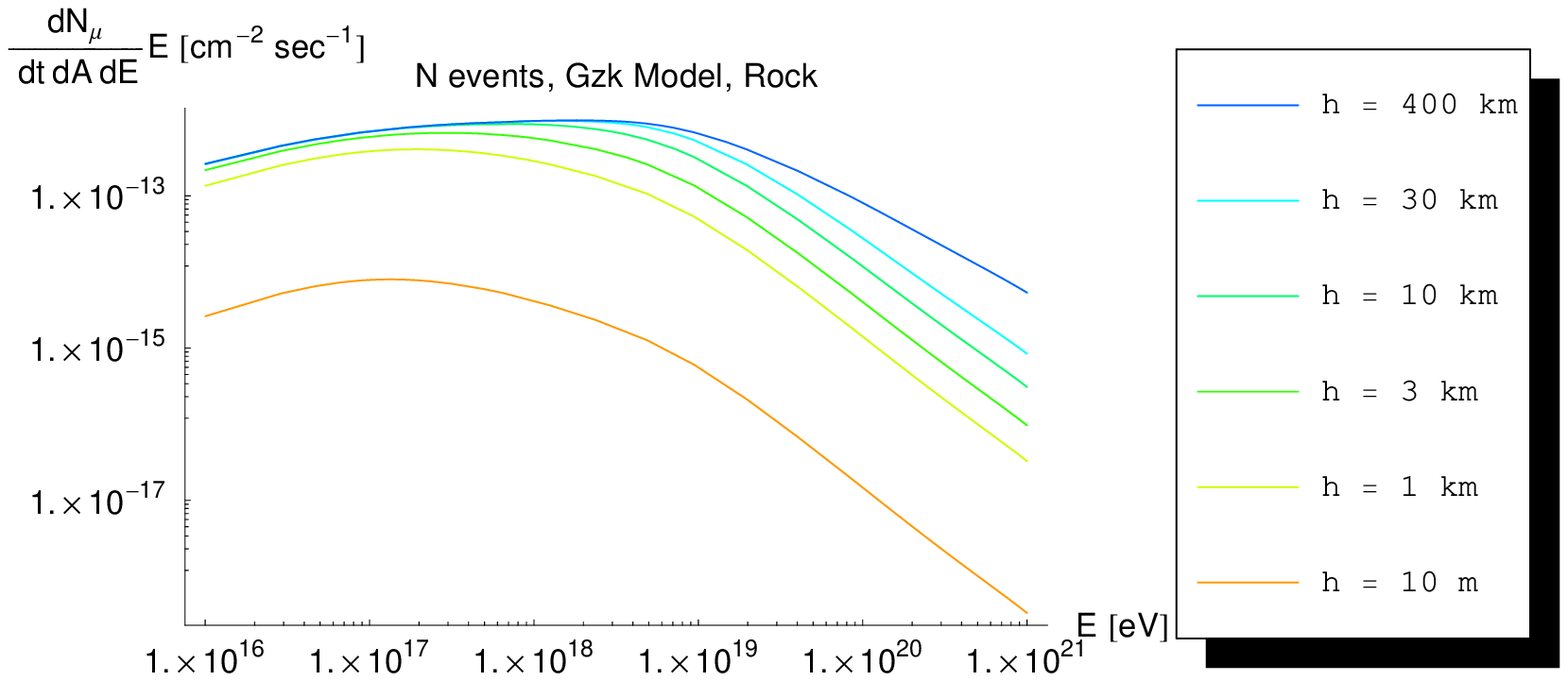}
\includegraphics[width=80mm,height=55mm,angle=0]{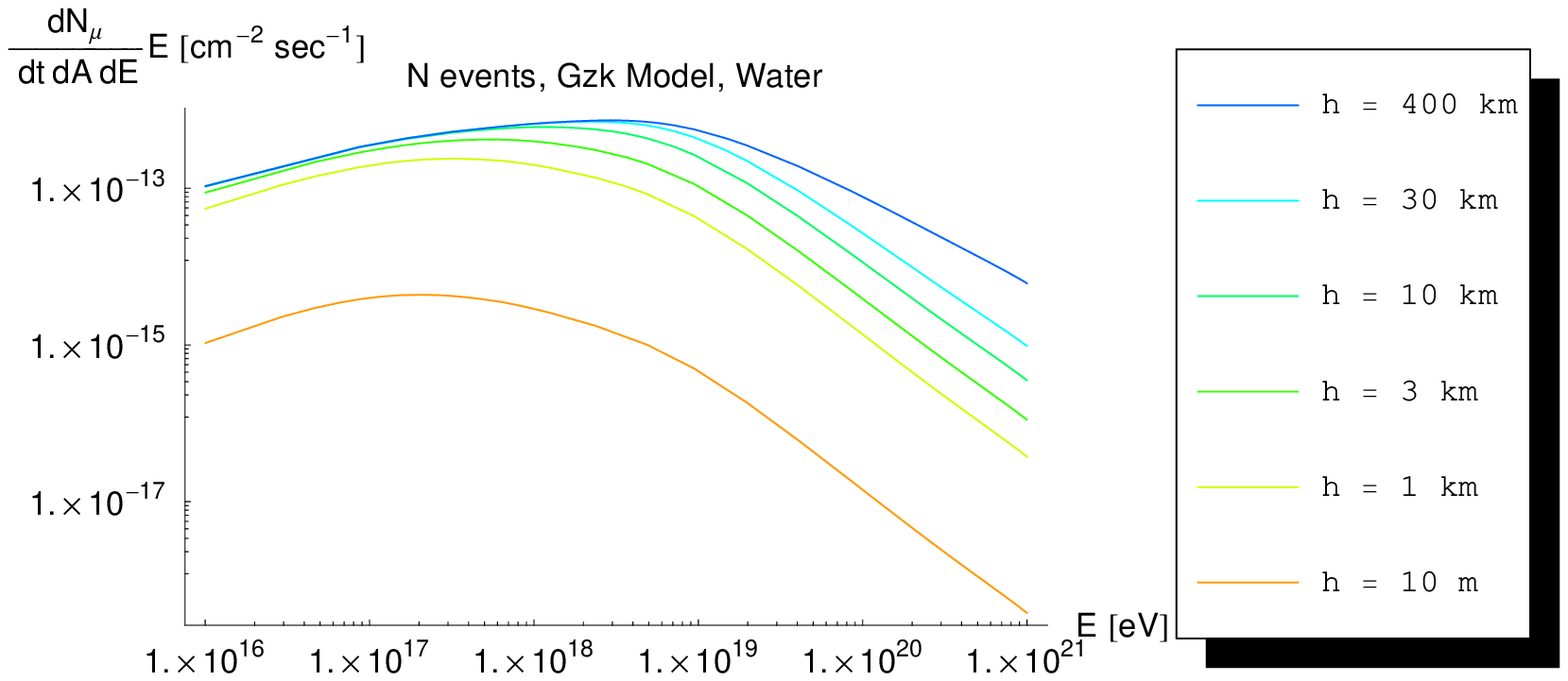}
\caption{{\em Left:}  The differential number flux (per unit area,
energy and time) of the expected number of events of $\mu$ leptons
from $\tau$ decay at different altitudes as a function of the
energy of the incoming neutrino $E_{\nu}$. Here we have assumed an
input GZK neutrino flux and an Earth outer layer made of {\em
rock}. {\em Right:} The differential rate of the expected number
of events of $\mu$ leptons from $\tau$ decay at different
altitudes as a function of the energy of the incoming neutrino
$E_{\nu}$. Here we have assumed an input GZK neutrino flux and an
Earth outer layer made of {\em water} The higher the observatory
the larger  the flux, and for $h$ = 400 km it can even exceed the
"noise" due to the atmospheric upgoing muons ($\Phi_{\mu, \, atm}$
$\sim 2 \times 10^{-13}$ cm$^{-2}$ s$^{-1}$ sr$^{-1}$).}
\label{differential}
\end{figure}

\begin{figure}[t] 
\includegraphics[width=80mm,height=65mm,angle=0]{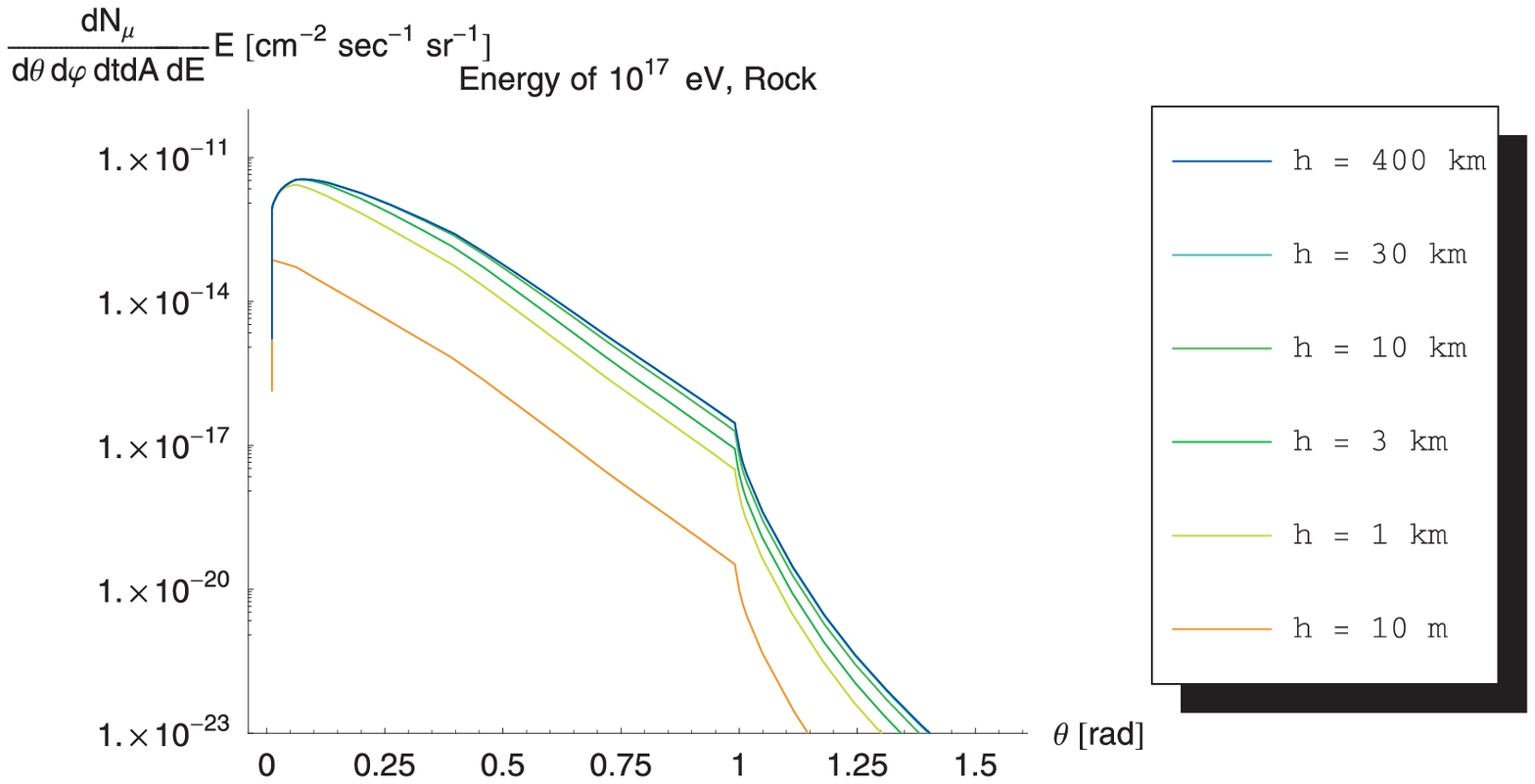}
\includegraphics[width=80mm,height=65mm,angle=0]{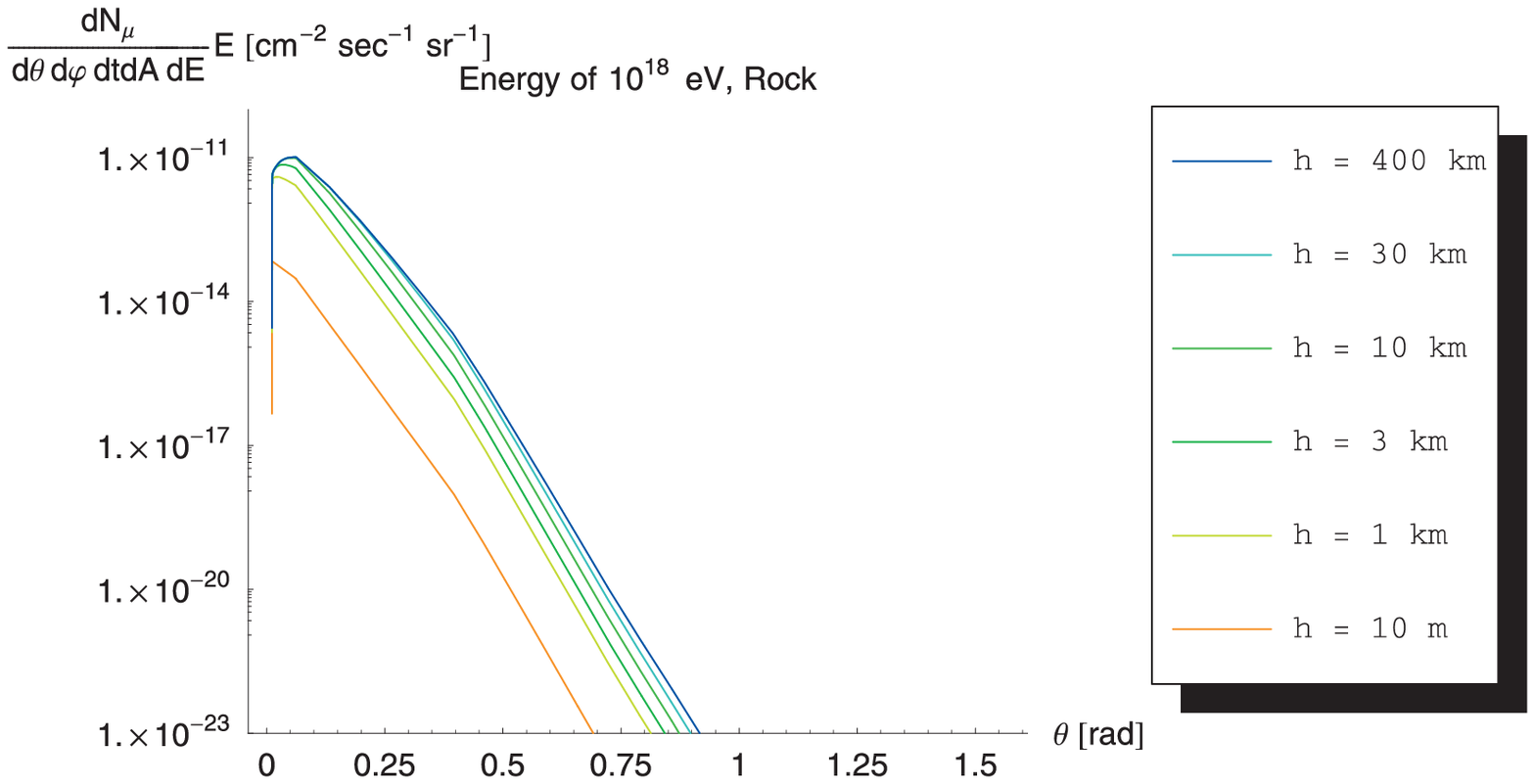}
\caption{The average differential number flux (per unit area,
solid angle and time)  of secondary muons from HorTaus at
different altitudes as a function of the aperture $\theta$ which
describes the line of sight below the horizon (i.e. $\theta =
\theta_{zenith} - 90^{\circ}$). The flux has been calculated for
two values of the energy of the $\tau$: $E_{\tau} = 10^{17}$ eV
({\em left panel}) and $E_{\tau} = 10^{18}$ eV ({\em right
panel}). We have assumed an input GZK-like neutrino flux and an
Earth outer layer made of rock. Note the discontinuity of the
angular spectrum in the $left$ panel, due to the sharp density
contrast  of the Earth at $\theta \simeq 1$ rad and the asymptotic
behaviour for $h \gg 1$ km. Because the Earth is nearly opaque to
EeV \nutau, only Earth-skimming neutrinos nearly horizontally are
visible. The discontinuity in the spectrum at $\theta \sim$ 1 rad
occurs also at energies as large as 10$^{18}$ eV ($right$ panel),
but at a lower flux, thus it is not included in this figure.}
\label{MU_FLUX17}
\end{figure}

\end{document}